\documentstyle[12pt,epsfig]{article}

\def\NPB{{\em Nucl. Phys.} B}
\def\PLB{{\em Phys. Lett.}  B}
\def\PRL{{\em Phys. Rev. Lett.}}
\def\PRD{{\em Phys. Rev.} D}

\def\be{\begin{equation}}
\def\ee{\end{equation}}
\def\bea{\begin{eqnarray}}
\def\eea{\end{eqnarray}}
\def\L{{\cal L}}

\def\eframe{\hbox{$\underline{E\ }\!\!|$\ }}
\def\half{\hbox{\small $\frac{1}{2}$}}

\begin{document}

\title{STRING COSMOLOGY: AN UPDATE}
\author{RAM BRUSTEIN \\
{\small \em Department of Physics, Ben-Gurion University,  
84105 Beer-Sheva, Israel }\\
{\small \tt e-mail:ramyb@bgumail.bgu.ac.il}}

\date{ }
\maketitle
{\vspace{-3in}\rightline{BGU-PH-98/01}

\vspace{3in}

\abstract{A string cosmology scenario (``pre-big-bang") postulates that the evolution of 
the Universe starts from a state of very small curvature and coupling,
undergoes a long phase of dilaton-driven kinetic inflation and at some later
time joins smoothly standard  radiation dominated cosmological evolution, thus
giving rise to a singularity free inflationary cosmology. I report on recent
progress in understanding some outstanding issues such as initial conditions,
graceful exit transition and generation of inhomogeneity perturbations.} 
\vfill 
\hrule
\vspace{.5in}
\noindent
Contribution to the proceedings of COSMO-97, International Workshop on
Particle Physics and The Early Universe, 15-19 September 1997, Ambleside, 
Lake District, England.

\newpage
\section{Introduction}

String theory predicts gravitation, but the 
gravitation it predicts is not that of standard general relativity. 
In addition to the metric, string gravity also contains a scalar dilaton,
that controls the strength of coupling parameters. 
An inflationary scenario \cite{gv1,gv2}, is based on the fact that
cosmological solutions to string dilaton-gravity 
come in duality-related pairs, an inflationary branch in which the Hubble 
parameter increases with
time and a decelerated branch that can be connected smoothly to a standard
Friedmann$-$Robertson$-$Walker expansion of the Universe with
constant dilaton. The scenario (``pre-big-bang") is that the
evolution of the Universe starts from a state of very small curvature and
coupling and then undergoes a long phase of dilaton-driven kinetic inflation.
Then, after spending some time in a high-curvature ``string phase" \cite{bggv}, 
a graceful exit transition occurs and the evolution joins smoothly standard 
radiation dominated cosmology, thus giving rise to a singularity
free inflationary cosmology. 

The two duality-related branches of lowest order solutions are singular. The
inflationary branch has a singularity in the future and the decelerated branch
has a singularity in the past, therefore, they cannot be connected
smoothly to form a single cosmology. However, because the Universe evolves
towards higher curvatures and stronger coupling, there will be some time when
the lowest order effective action can no longer reliably describe the dynamics
and it must be corrected. Corrections  come
from two sources. The first are classical corrections, due to the finite size of
strings, arising when the fields are varying over the string length scale
$\lambda_s=\sqrt{\alpha'}$. These terms are important in the regime of large
curvature.  The second are quantum loop corrections. The loop expansion is
parameterized by powers of the string coupling parameter $e^\phi=g_{string}^2$,
which in the models that we consider is time dependent.  So quantum corrections
will become important when the dilaton $\phi$  becomes large. As we will see,
the two types of corrections can induce a graceful exit transition. Even in the
absence of a detailed description of the high curvature string phase, assuming
that a graceful exit transition did occur, it is possible to find
experimentally observable consequences of string cosmology models.

Most of the experimentally observable consequences of  string
cosmology models based on the pre-big-bang scenario arise from the 
inhomogeneity perturbations generated around the
homogeneous and isotropic background. Inhomogeneities get produced 
during the dilaton-driven phase
 by the standard mechanism of amplification of vacuum fluctuations 
\cite{birrdav,gv2,pert}. Deviations from homogeneity and isotropy are generated by 
quantum  fluctuations around the homogeneous and isotropic background and then
amplified by the accelerated expansion of the Universe.
In addition to the dilaton and metric, string theory contains 
many other fields that have trivial expectation values  and do not affect the
classical solutions,  but they do fluctuate. 

\section {Initial Conditions}

Initial conditions are an essential part of any cosmology, as stressed by 
Hawking\cite{inhawk}. Since the simplest dilaton-driven inflationary solution has a
future singularity, and is completely regular in the past, it might have  given
the wrong impression that the evolution could be extended backwards in time
indefinitely. We now know that this is only correct for some strictly
homogeneous and  isotropic solutions.  The  question
``how did the Universe find itself in the dilaton-driven inflationary phase" is a
valid question, although it may well be that its resolution lies outside of
physics.

In all inflationary models, to solve the problems of standard cosmology, the
initial conditions have to be rather special.  For the simplests string
cosmology  models the Hubble parameter $H$, of the 
inflationary branch satisfies $H=\frac{1}{\sqrt{3}} (t_0-t)^{-1}$ and
$\dot\phi= (1+\sqrt{3})(t_0-t)^{-1}$ for $t<t_0$. The  number of
e-folds from $t_i$ until $t_f$ is $Z= \frac{a(t_f)H(t_f)}{a(t_i) H(t_i)}=
e^{1/\sqrt{3}\left(\phi(t_f)-\phi(t_i)\right)}$. To ensure at least 60 e-folds 
of inflation during the dilaton-driven phase, the
dilaton has to start such that the coupling $e^{\phi/2}$ is very small and
remains  perturbative throughout the evolution. 

The basic postulate of the scenario is that the initial curvature
and the initial coupling are very small,  however this does not necessarily
mean that the Universe
starts in a completely homogeneous and isotropic state.
The issue of initial conditions for the dilaton-driven phase has been
investigated by several authors \cite{inhom1}-\cite{inhom4}  and we
review the results. We introduce the problem by looking at the equation for $H$ in the
presence of spatial curvature to model the evolution of under- and over-dense
regions, $\dot H= H\sqrt{3H^2-{6k}/{a^2}}-2k/a^2$,  where $k=0,\pm 1$ determines
the spatial curvature. Even without solving the equation we see that the flat
($k=0$) inflationary solution is an attractor, since the contribution of the
curvature term decreases rapidly as the Universe expands. Running the equation
backwards in time we see that the effects of curvature increase and the solution
deviates substantially from the dilaton-driven solution. 

Effects of inhomogeneities  were first discussed in \cite{inhom1}.  It
was found that the inflationary branch is an attractor for a class of
inhomogeneous initial conditions and that  a range of initial conditions
 will lead to the inflationary branch. The conclusion of \cite{inhom2}  is
that the results of \cite{inhom1} are correct, but the onset of inflation is
delayed by initial homogeneities and curvature and therefore the requirement on
the initial value of the dilaton and curvature are more severe depending on the
initial curvature and inhomogeneities.
The conclusion of \cite{inhom4} is that  the results of  \cite{inhom2}
are correct,  but there are two classical moduli of string theory, the initial
scale factor and initial coupling that are free and there is always a choice
that will guarantee a long duration of inflation, 
\begin{equation}
Z\sim\min\left[e^{1/\sqrt{3}\left(\phi(t_f)-\phi(t_i)\right)},
\left(\lambda_s^2 {\cal R}(t_i)\right)^{-\frac{\sqrt{3}+1}{2 \sqrt{3}}}\right],
\end{equation}
where ${\cal R}(t_i)$ is the initial curvature. That the initial coupling has to
be small enough, we have already seen, in addition, the initial curvature has to
be small enough so that it does not become too big too early and disrupt the
dilaton-driven inflation. The conclusion of \cite{inhom3} is that the
results of  \cite{inhom2} are correct but that in estimating the total
number of e-folds, 
 the expansion during the string phase should also be taken into account and
therefore the constraints on the initial conditions should be relaxed.

The results show, in my opinion, that as in all
models of inflation, initial conditions do need to be specified, 
and that they are rather special. To decide quantitatively how likely or unlikely
they are, a theory of initial conditions which does not exist at the 
moment has to be developed.

\section {Graceful Exit }
 
The two duality related branches of solutions of the lowest order string 
effective equations are separated by a singularity. Additional fields or 
correction terms  need to be added to make a smooth ``graceful exit'' 
transition between branches possible. 
In \cite{bv,kmo} it was shown that the transition is forbidden  
for a large class of fields and potentials, a result which was reinforced by
many subsequent investigations \cite{too}. In \cite{BM} we proposed to use 
an effective
description in terms  of sources that represent arbitrary corrections to the
lowest order  equations and were able to formulate a set of necessary 
conditions for graceful exit and to relate them to energy conditions
appearing in singularity theorems of general relativity 
\cite{he}. 
We showed that a successful exit requires violations of the
null energy condition (NEC) and that this violation is associated with
the change from a contracting to an expanding universe (bounce) in the 
``lowest order Einstein frame'' (\eframe), defined by a conformal change of variables.
Since most classical sources obey NEC this
conclusion hints that quantum effects, 
known to violate NEC in some cases, may be
the correct sources to look at.  

The ${\scriptstyle 00}$ equation of motion is quadratic and may be conveniently
written in the string frame, 
\begin{equation} 
\dot\phi=3 H \pm \sqrt{3 H^2+e^{\phi} \rho},
\label{rhoeq}
\end{equation} 
where we have fixed  our 
units such that $16\pi\alpha'=1$. The choice of sign here corresponds
to our designation of (+) and ($-$) branches.  

In \cite{BM2}, relying on a demonstration that classical corrections
can limit curvature \cite{gmv}, we were able to find an explicit model that
satisfies  all the necessary conditions and to produce the first example of a 
complete exit transition. 

The explicit correction Lagrangian we use to produce the concrete example presented in
Fig. 1 is  $\half \L_{c}=e^{-\phi}
(\frac{R_{GB}^2}{4}- \frac{(\nabla \phi)^4}{4}) -
1000 (\nabla \phi)^4 + 1000 e^{\phi} (\nabla \phi)^4.$ 
The first term is in the form of $\alpha'$ corrections
examined in \cite{gmv}, 
the second and third are plausible forms for the
one and two loops corrections respectively.  The large coefficients 
account for the expected large number of degrees of freedom contributing
to the loop. The signs of these terms are deliberately chosen to force
the exit.
We have checked that qualitatively similar evolution is obtained for a 
range of coefficients, of which the evolution in Fig. 1 is a representative.

\begin{figure}[h]
\begin{center}
\vspace{.2in}
\epsfig{file=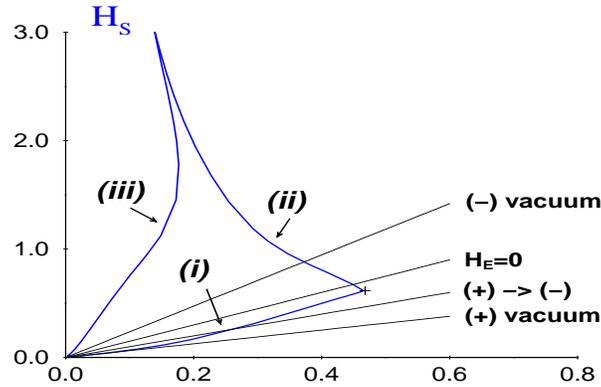,width=8cm,height=4cm,bbllx=12,bblly=60,
 bburx=576,bbury=430}
\end{center} 
\caption{\small\it A model of a complete graceful exit.} 
\end{figure}

We set up initial conditions in
weak coupling near the (+) branch vacuum and a numerical integration yields
the evolution shown in Fig. 1  in the $\dot \phi$,$H_S$ phase space.
We have also plotted lines marking important landmarks in the evolution,
the (+) and ($-$) vacuum ($\rho=0$ in (\ref{rhoeq})), the line of branch 
change $(+) \rightarrow (-)$ (square root vanishing in (\ref{rhoeq})) 
and the position
of the \eframe bounce ($H_E=0$, $\dot \phi=2 H_S$). 

The evolution falls into distinct phases.
{\bf Phase (i)}
This is the inflationary phase. As curvature becomes large we see deviation
induced by the $\alpha'$ corrections, and without
influence from other corrections the solution would settle into the fixed 
point noted in \cite{gmv}, marked with a '+'. 
The solution does cross 
the line of branch change, but does not execute the \eframe bounce 
required by a complete exit, corresponding to the fact it does not 
violate NEC in the \eframe.
{\bf Phase (ii)}
Eventually the loop corrections become important, while the coupling is still
perturbative because of the large correction coefficient. The first to do so is
the one loop correction. Since we require further NEC violation to complete the
\eframe bounce, we have chosen the sign of the one loop correction to provide
this  and in this phase corrections are dominated by this term. As a result a 
 bounce occurs and the evolution proceeds into the $\rho>0$ region. 
We checked that other forms of loop correction will have the same effect
if they are introduced with a coefficient allowing NEC violation. Without
further corrections this solution would continue to grow into regions
of larger curvature and stronger coupling. We refer to this era as
``correction dominated'' and we also find there are obstacles to 
stabilizing the dilaton with standard mechanisms like capture in a 
potential or radiation production.
{\bf Phase (iii)}
To offset the destabilizing NEC violation we have introduced the 
two loop correction with the opposite sign, allowing it to overturn
the NEC violation when it becomes dominant as $\phi$ continues to grow.
Indeed during this phase we see the expansion decelerating, dilaton 
growth stabilizing, and the corrections vanishing.
We have also checked that in this phase the dilaton can be 
captured into a potential minimum or halted by radiation production.
This phase can be smoothly joined to standard cosmologies.

\section {Inhomogeneity perturbations and particle production}

Until recently, most of the  produced particle spectra were
found to rise sharply with frequency, a property with interesting
consequences \cite{bggv,gravitons,photons,dilatons}, but also some
disadvantages. In particular, at very large wavelengths there is almost no
power, making these inhomogeneity perturbations an unlikely source for 
large scale anisotropy. The standard explanation given for the generic
spectral frequency dependence is that since the curvature increases sharply,
particle production also increases and hence the resulting spectrum. However, 
we now know that this standard explanation is not always correct. The first
hint came from axionic vacuum perturbations \cite{axions} which 
exhibited a decreasing spectrum. 
Although sharply rising spectra are indeed common,  flatter
spectra or even decreasing spectra are just as likely \cite{bh,bmuv}. 
In addition to the obvious dependence on the background solution,
the spectrum depends also on the spin of the particle, the type of dilaton
prefactor, the coupling of the particle to internal moduli, and whether 
it is massless or massive, revealing a  rich range of spectral shapes, 
of which many more deserve further individual attention.

The action for each field's perturbation is obtained by expanding the 4
dimensional effective action of strings, which generically, for a tensor field
of rank $N$, has the form  
$
{\small\frac{1}{2}} \int d^4 x \sqrt{-g}
e^{l\phi}{\cal L}^{(2)} (T_{\mu_1,...,\mu_N})
$, where the parameter $l$ in the
dilaton prefactor is determined by the type of field.  Setting the dilaton and
metric at their background  values, $g_{\mu\nu}(\eta)=a^2(\eta )\eta_{\mu\nu}$, 
$\phi(\eta)$ (where $\eta$ is conformal time) results in a
 quadratic action for each physical component of the perturbation $\psi$,  
\begin{equation} 
A={\small\frac{1}{2}} \int d\eta\ d^3x\ S^2(\eta)
\left( \psi ^{\prime 2}- (\nabla \psi )^2-M^2 a^2 \psi ^2\right)    
\label{pertact}  
\end{equation} 
where $'$ denotes $\partial/\partial\eta$ and $M$ is the mass of the perturbed
field. The function $S(\eta)$ (equal to $a^{m}e^{l\phi/2 }$ for the simplest 
case but can be
quite complicated in general) is the ``pumping-field". 

The linearized equation of motion, satisfied by the  Fourier modes of the 
field perturbation $\psi$,
derived from the action (\ref{pertact}) is the following 
\begin{equation}
\psi_k^{\prime \prime}+2\frac{S'}{S} \psi_k ^{\prime }
+\left(k^2+M^2a^2\right) \psi_k =0.  
\label{perteq2}
\end{equation}

The general solution of the perturbation equation is a linear combination of
two modes, one which is approximately constant outside the horizon and one
which is generically time dependent outside the horizon. We understand the
appearance of a constant mode as the freezing of the perturbation amplitude,
since local physics is no longer active on such scales. The  existence of the
time dependent mode can be most easily understood in terms of a constant mode
of the conjugate momentum of $\psi$, $\Pi=S^2(\eta)\psi'$. The amplitude of
the conjugate momentum also freezes outside the horizon,  since local
physics is no longer active on such scales. This forces $S^2(\eta)\psi'$ to be
approximately constant leading to a ``kinematical" time dependence of 
$\psi$ \cite{sbarduality}.
The number density and energy density of the produced particles can be easily
read off from the solutions of eq.(\ref{perteq2}) in a standard way
\cite{birrdav}. The function $S(\eta)$ determines the resulting spectrum of
produced inhomogeneities (or equivalently, particles) 
and may depend in a non-trivial way on
$a(\eta)$, $\phi(\eta)$, and additional variables. If the background
solution changes, so does $S$, even if its functional dependence on the dilaton
and scale factor remains the same. Therefore, a rich array of spectral indices
appears, as the example of Table 1. shows. For a comprehensive list of
possibilities, and  detailed explanation see \cite{bh,bmuv}.
\begin{table}[t]
\caption{\small\it 
A variety of spectral indices for  different fields, assuming the simplest
background evolution. For internal axions a range of possible internal backgrounds
was considered.
The spectral index $n$
determines the energy density spectrum ($n=1$ corresponds to a flat spectrum). 
}
\label{tab:exp} \begin{center}
\begin{tabular}{|c|c|}
\hline
 field & spectral index $n$  \\ \hline
gravitons, dilatons  & 4 \\ \hline
model-independent axions & $0.54$ \\ \hline
internal axions  & from 0 to 4 \\ \hline
RR axions  & 2.26 \\ \hline
heterotic gauge bosons  & $3.27$ 
\\ \hline
\end{tabular}
\end{center}
\end{table}

In \cite{bh} we solved the perturbation equation by first solving the early
time equation with boundary conditions of normalized vacuum fluctuations. We
assumed that both the constant mode of $\psi$ and the constant mode of $\Pi$
remain constant while outside the horizon during the string phase, thus
bridging the gap of unknown background evolution during the string phase, at the
expense of introducing only two string phase parameters. We then matched the
early time solutions to the late time solutions. In \cite{bmuv}, and in
previous calculations, an explicit background solution during the string phase
was assumed. In all cases where a comparison is possible the results using the
two methods agree.

\section*{Acknowledgment}
Research supported in part by the Israel Science Foundation administered
by the Israel Academy of Sciences and Humanities. I wish to thank
R. Madden for useful discussions and comments about this update
which is based mostly on joint work with him and with M. Hadad. I also wish
to thank A. Buonanno, K. Meissner, C. Ungarelli and G.
Veneziano for discussions about their work.

\end{document}